\documentclass{article}
\usepackage[a4paper, margin=1.3in]{geometry}
\usepackage{listings}
\usepackage{tabularx}
\usepackage{float}
\usepackage[english]{babel}
\usepackage[utf8]{inputenc}
\usepackage{amsmath}
\usepackage{csquotes}
\usepackage[authoryear]{natbib}
\usepackage{graphicx} 

\title{Revisiting John Snow's Cholera Map: A Data Visualisation Case Study for Statistical Education}
\author{Niamh Mimnagh}
\date{}
\begin{document}

\maketitle


\section{Abstract}
Data visualisation is a fundamental tool in statistical analysis, enabling the identification of patterns and relationships that might otherwise remain hidden in raw data. One of the most famous historical examples is John Snow's 1854 cholera map, which demonstrated the spatial clustering of cholera cases around a contaminated water pump in London. This study explores how Snow's visualisation can be effectively incorporated into statistics education as an interactive case study. Using R, we outline the steps involved in reproducing Snow's cholera map, demonstrating geospatial data manipulation, visualisation techniques, and spatial analysis. We discuss the pedagogical benefits of historical case studies in statistics courses, emphasising their role in fostering curiosity, critical thinking, and technical proficiency. Additionally, we explore how these methods can be extended beyond epidemiology to applications in public health, urban analytics and environmental science. By integrating historical datasets with modern computational tools, educators can create engaging, hands-on learning experiences that reinforce core statistical principles while illustrating the real-world impact of data analysis. 

\section{Introduction}
Data visualisation plays an important role in statistical analysis and decision-making, allowing patterns and relationships to emerge that might otherwise remain hidden in raw data. One of the most famous early examples of discovery driven by visualisation comes from John Snow's 1854 cholera map \citep{snow1856mode}, which is presented in Figure \ref{fig:map}, and is widely recognised as a pioneering work in epidemiology and spatial statistics \citep{tufte1997visual}. By plotting cholera cases on a street map of Soho, London, Snow succeeded in demonstrating a spatial pattern that linked cases to a contaminated water pump, providing critical evidence against the dominant miasma theory of disease transmission \citep{brody2000map}. His analysis contributed to the foundation of modern epidemiology and remains a compelling example of how statistical thinking, when combined with effective visualisation, can drive scientific progress \citep{friendly2001milestones}. 
\begin{figure}
    \centering
\includegraphics[width=0.6\linewidth]{snow_map.jpg}
    \caption{John Snow’s original 1854 cholera map, illustrating the clustering of cholera cases around the Broad Street pump.}
    \label{fig:map}
\end{figure}
Beyond its historical significance, John Snow's work serves as an excellent teaching tool for students learning about data visualisation, spatial analysis and statistical reasoning . The process of mapping the data, exploring patterns, and drawing inferences mirrors the methods used by statisticians and data scientists. In an era where computational tools such as R, Python and GIS software enable increasingly sophisticated analyses, revisiting Snow's approach using contemporary methods offers a unique opportunity to engage students with both historical and modern perspectives on data science. 

In this article, I explore how Snow's visualisation can be incorporated into statistics education as an interactive case study. Using R, I outline the steps involved in reproducing the cholera map, demonstrating how students can use geospatial data manipulation and visualisation techniques to uncover the same insights Snow did over 150 years ago. 

Snow's case study provides a foundation for discussing widely used statistical techniques. For instance, bar plots and histograms can illustrate disease counts across different city districts, scatter plots can visualise spatial clustering, and kernel density estimation can provide a modern alternative to Snow's hand-drawn Voronoi diagram. Regression modelling can extend the case study to infer relationships between environmental factors and disease spread.

I also discuss the pedagogical benefits of using historical case studies in statistics courses, emphasising how they can foster curiosity, critical thinking, and technical proficiency. Finally, I consider how this approach can be extended beyond epidemiology, illustrating broader applications in public health, urban analytics, and environmental science, where similar data visualisation and modelling techniques are vital.

By blending historical datasets with modern computation techniques, we can create engaging, hands-on learning experiences that not only reinforce core statistical principles, but also help students appreciate the real-world impact of data analysis. Snow's cholera map is more than  just a historical artifact, it is a powerful example of how data visualisation can reveal insights, shape scientific understanding, and drive evidence-based decision-making, making it an invaluable resource for a statistical education.

\section{The Educational Value of Data Visualisation}
Data visualisation is a fundamental tool in statistics education, allowing students to explore patterns, uncover relationships, and communicate insights effectively \citep{nolan2016teaching}. Unlike raw numerical summaries, visual representations provide an intuitive and immediate understanding of complex datasets. Whether through scatterplots, histograms or geospatial maps, visualisation helps students bridge the gap between abstract statistical concepts and real-world interpretation. 

John Snow's 1854 cholera map offers a compelling case study in this regard. By plotting cholera cases spatially, Snow was able to reveal an epidemiological pattern that was not immediately obvious from tabular data alone. This example underscores a key lesson for students: data visualisation is not just about making data look appealing, it is a vital step in the analytical process that can lead to new discoveries and enables more informed decision-making \citep{tukey1977exploratory, unwin2020data}.

In modern statistics and data science curricula, incorporating historical case studies like Snow's cholera map can serve multiple educational purposes:

\begin{enumerate}
    \item Enhancing Statistical Thinking
    
By visualising the spread of cholera cases, students engage in exploratory data analysis, a crucial step in any statistical workflow. This encourages pattern recognition, hypothesis generation and critical thinking, skills that are essential for informed decision-making. It also reinforces the idea that data should be represented visually before any formal modelling begins, as unexpected insights may emerge \citep{tukey1972some}.
    \item Introducing Spatial Data and Geographic Information Systems

Traditional statistics courses often focus on numerical and categorical data, but spatial data introduces new dimensions of analysis. Snow's map allows students to work with coordinate-based data, spatial clustering techniques, and geospatial visualisation tools in R \citep{unwin2018graphical}. It serves as an introduction to Geographic Information Systems (GIS) \citep{cromley2011gis} concepts, which are widely used in epidemiology, urban planning, and environmental science.
    \item Connecting Historical Data to Modern Applications

Many of the same principles Snow applied in the $19^{\text{th}}$ Century as used today in public health monitoring, pandemic modelling and risk assessment. Students can draw parallels between Snow's cholera outbreak investigation and modern epidemiological studies, such as COVID-19 hotspot mapping \citep{franch2020spatial, zhang2020spatial} or air pollution analysis \citep{brunekreef2002air}. This connection reinforces the relevance of statistical reasoning beyond the classroom.
    \item{Encouraging Computational Skill Development}
    
Implementing Snow's visualisation in R introduces students to essential programming concepts, including data wrangling (e.g. dplyr \citep{dplyr} and tidyverse \citep{tidyverse}), geospatial mapping (e.g. ggplot2 \citep{ggplot2} and sf \citep{sf1, sf2}), and reproducible research practices. This hands-on experience helps student build confidence in working with real-world data and computational tools. It demonstrates how open-source software can be used for both historical data analysis and contemporary statistical applications
\end{enumerate}

By integrating historical data visualisation projects into statistics education, instructors can create engaging, interdisciplinary learning experiences that blend historical inquiry, data science and statistical reasoning. John Snow's cholera map is not just an artifact of epidemiological history, it is a timeless example of the power of data visualisation as an investigative and educational tool.

\section{Recreating John Snow's Map Using R}
Reconstructing John Snow's 1854 cholera map using modern statistical tools provides an engaging hands-on exercise for students learning data visualisation, geospatial analysis and epidemiological modelling. With R, we can replicate Snow's approach by using real historical data, reinforcing key statistical concepts while demonstrating the way computation tools enhance our ability to analyse and visualise spatial data. 

In this section, I outline the key steps involved in reproducing Snow's map using R, focusing on data acquisition, spatial visualisation and interpretation.

\subsection{Preparing the Data}

To recreate the cholera map, we first need data on:
\begin{itemize}
    \item Cholera cases: the locations where cholera deaths were recorded.
    \item Water pumps: The suspected sources of contamination.
    \item Street layout: the geographical context for visualisation.
\end{itemize}

These datasets are available from historical sources and have been digitised for analysis in R. The {SnowData} package \citep{snowdata} provides a clean version of Snow's data, which can be loaded as follows:

\begin{verbatim}
# Load required packages
library(ggplot2)
library(sf)       # For spatial data
library(dplyr)    # For data wrangling
library(SnowData) # Contains Snow's cholera data

# Load the cholera cases dataset
data(cholera_cases)
data(pump_locations)

\end{verbatim}

At this stage, we have two key spatial datasets: one containing the cholera deaths and another with the water pump locations.

\subsection{Mapping Cholera Cases and Water Pumps}

Next, we visualise the data using {ggplot2}, overlaying the cholera cases and water pumps onto a simple map.

\begin{verbatim}
# Plot cholera deaths and water pumps
ggplot() +
  geom_point(data = cholera_cases, aes(x = Easting, y = Northing), 
            color = "red", alpha = 0.6, size = 2) +
  geom_point(data = pump_locations, aes(x = Easting, y = Northing), 
            color = "blue", size = 4, shape = 17) +
  labs(title = "John Snow's 1854 Cholera Map",
       subtitle = "Red dots = Cholera deaths, Blue triangles = Water pumps") +
  theme_minimal()
\end{verbatim}

\begin{figure}[H]
    \centering
    \includegraphics[width=\linewidth]{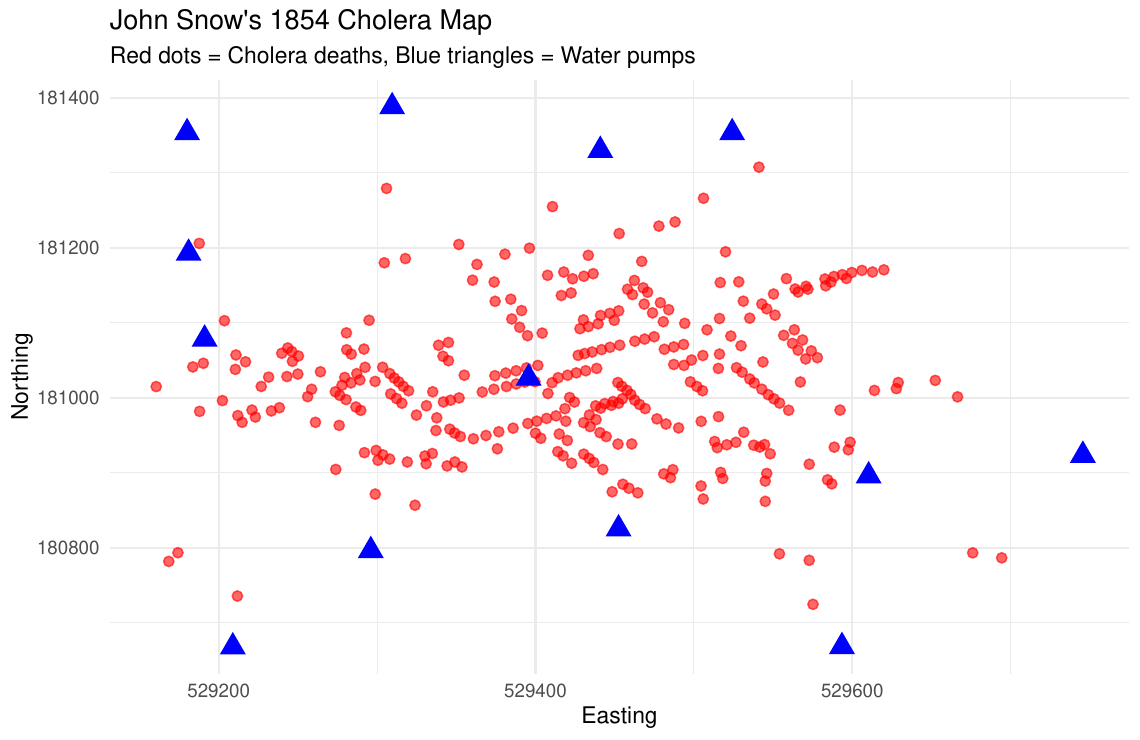}
    \caption{Distribution of cholera cases (red dots) and water pumps (blue triangles) during the 1854 Broad Street outbreak.}
    \label{fig:plot1}
\end{figure}

This initial visualisation highlights the spatial clustering of cholera cases around certain water pumps, reflecting Snow's original analysis.

\subsection{Adding Street Layout for Context}
To make the visualisation more informative, we can overlay the data onto the original Soho street map. This can be achieved using a digitised version of Snow's map, available as part of the {SnowData} package, or OpenStreetMap data.

\begin{verbatim}
library(sf)
library(osmdata)
library(ggplot2)

# SnowData uses the British National Grid (EPSG:27700), 
# so we define the bounding box using the same projection

soho_bbox_27700 <- st_bbox(c(xmin = 529150.0, 
                             ymin = 180720.6, 
                             xmax = 529750.9, 
                             ymax = 181370.5), 
                           crs = st_crs(27700))

# Convert to EPSG:4326 for OSM request
soho_bbox_4326 <- st_bbox(st_transform(st_as_sfc(soho_bbox_27700), crs = 4326))

# Request OSM data using the transformed bbox
streets <- opq(bbox = soho_bbox_4326) %>%
  add_osm_feature(key = "highway") %>%
  osmdata_sf()

# Convert OSM data to EPSG:27700
streets <- st_transform(streets$osm_lines, crs = 27700)

# Plot in EPSG:27700 using ggplot2
ggplot() +
geom_sf(data = streets, color = "gray50", size = 0.5)+
geom_point(data = cholera_cases, 
            aes(x = Easting, y = Northing), 
            color = "red", alpha = 0.6, size = 2) +
geom_point(data = pump_locations, 
            aes(x = Easting, y = Northing), 
            color = "blue", size = 4, shape = 17) +
theme(panel.grid = element_blank(),    
        axis.title.x = element_blank(), 
        axis.title.y = element_blank(),
        panel.background = element_rect(fill = "white"), 
        plot.background = element_rect(fill = "white")) +
coord_sf(crs = 27700, datum = st_crs(27700)) 
\end{verbatim}

\begin{figure}[H]
    \centering
    \includegraphics[width=\linewidth]{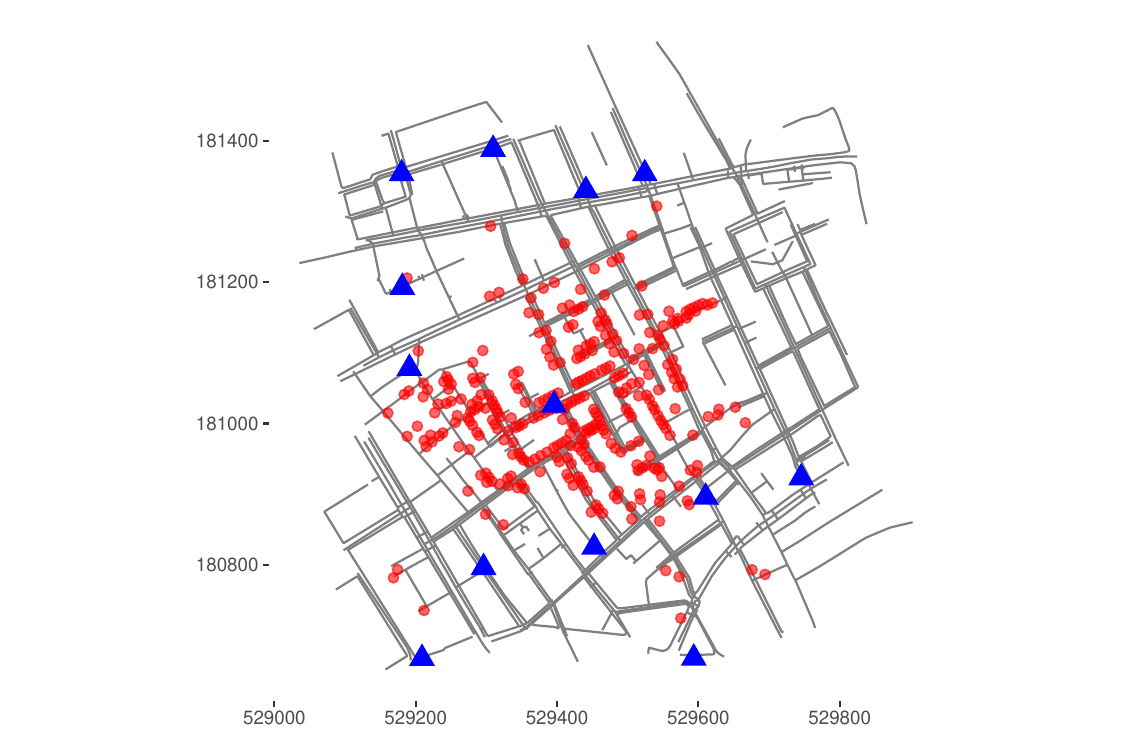}
    \caption{Cholera cases (red dots) and water pumps (blue triangles) overlaid on the Soho street network during the 1854 Broad Street outbreak. The spatial pattern highlights the proximity of cases to specific pumps.}
    \label{fig:plot2}
\end{figure}

Alternatively, as a more complicated exercise, students may attempt to accurately recreate John Snow's original 1854 map. Students may begin by loading the georeferenced map to use as a background. 

\begin{verbatim}
map <- load_map()
data(cholera_cases)
data(pump_locations)
\end{verbatim}

John Snow's original map represents the cholera outbreak in the Soho district of London, showing the locations of cholera deaths as black bars stacked along the streets. In order to create this plot, students must generate these bars, and rotate them accordingly. This can be done using a user-defined function, and Angle and Count columns in the \texttt{cholera\_cases} dataset, to specify the necessary angle of rotation and the height of the black bars, respectively.

\begin{verbatim}  
rotate_rectangle <- function(x, y, count, angle) {
 # Convert angle from degrees to radians
 theta <- angle * pi / 180
  
# Fixed width and height controlled by count
 hw <- 2      # Half of fixed width 
 hh <- count  # Height is proportional to count 
  
 # Original rectangle corners relative to base point
 corners <- matrix(c(-hw, 0,    # Bottom-left corner
                     hw, 0,     # Bottom-right corner
                     hw,  hh,   # Top-right corner
                     -hw,  hh), # Top-left corner
                     ncol = 2, byrow = TRUE)
  
 # Rotation matrix
 rotation_matrix <- matrix(c(cos(theta), -sin(theta), 
                            sin(theta),  cos(theta)), 
                            ncol = 2)
  
 # Rotate corners and translate to base point (X, Y)
 rotated_corners <- corners %*% rotation_matrix
 rotated_corners <- data.frame(rotated_corners)
 colnames(rotated_corners) <- c("x", "y")
 rotated_corners$x <- rotated_corners$x + x
 rotated_corners$y <- rotated_corners$y + y
 return(rotated_corners)
}

# Generate rotated rectangles
rectangles <- cholera_cases %>%
  rowwise() %>%
  mutate(corners = list(rotate_rectangle(Easting, Northing, Count*2, Angle))) %>%
  unnest(corners) %>%
  ungroup() %>%
  mutate(id = rep(1:nrow(cholera_cases), each = 4)) 
\end{verbatim}

\begin{verbatim}
# Map and overlay rotated rectangles
gplot(map, maxpixels = 5e6) + 
  geom_tile(aes(fill = value)) +
  scale_fill_gradient(low = 'black', high = 'white')+
  geom_point(data = pump_locations,
            aes(x = Easting, y = Northing), 
            colour = "blue") +
  geom_polygon(data = rectangles,
        aes(x = x, y = y, group = id),
        color = "black") +
  theme(axis.text.x = element_blank(),
        axis.title.x = element_blank(),
        axis.text.y = element_blank(),
        axis.title.y = element_blank(),
        axis.line = element_blank(), 
        legend.position="none") 
         
\end{verbatim}
\begin{figure}[H]
    \centering
    \includegraphics[width=\linewidth]{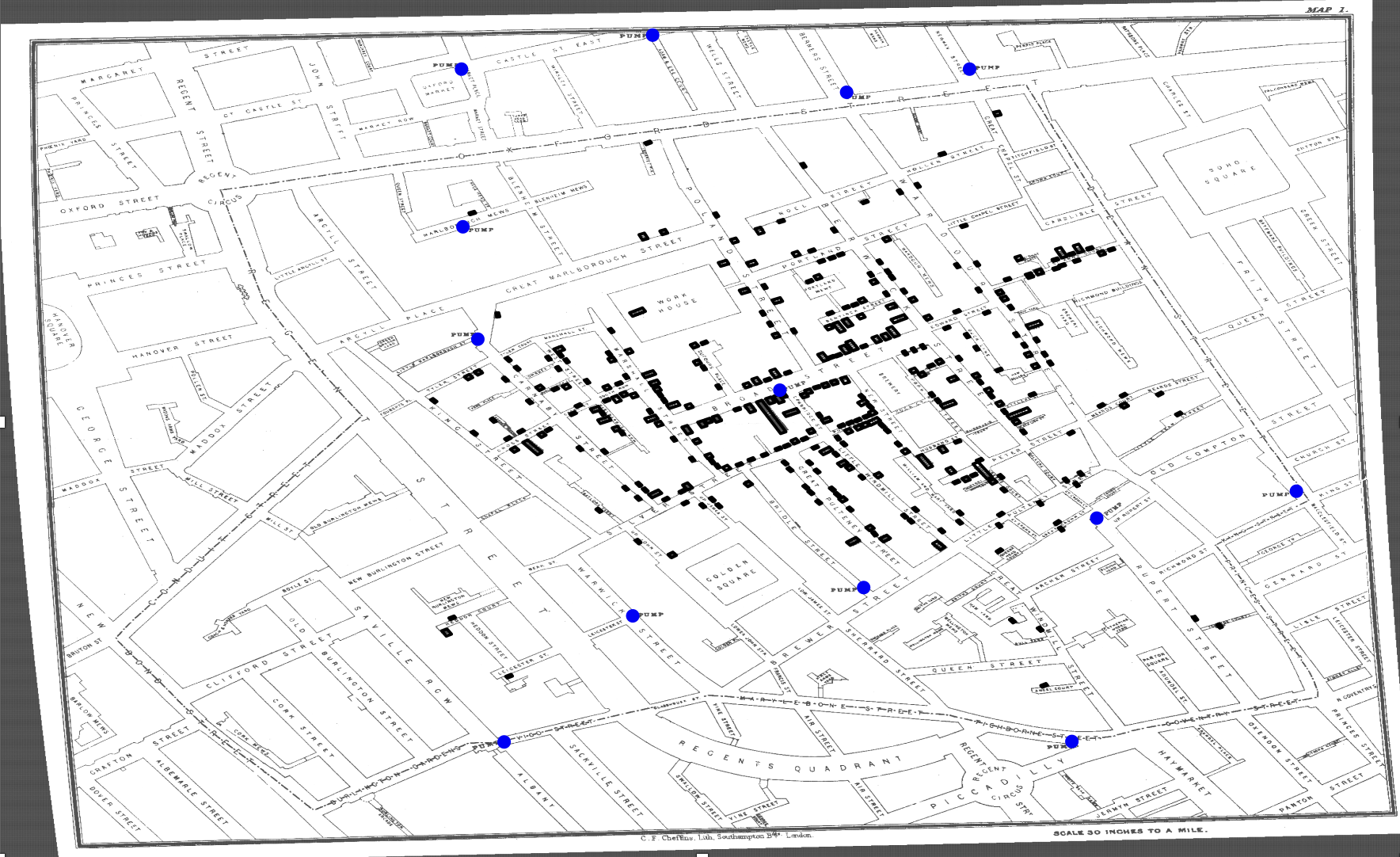}
    \caption{Cholera cases (black bars) and water pumps (blue dots) overlaid on John Snow's original 1854 map of the Soho outbreak. This visualisation replicates Snow's historic analysis linking cholera to contaminated water sources.}
    \label{fig:plot3}
\end{figure}
The resulting map is slightly warped due to stretching that occurred in order to match the historical map to modern geographic coordinates during georeferencing. However, this approach provides students with a historically accurate and visually compelling representation of Snow's findings, reinforcing the importance of geospatial analysis in epidemiology.

\subsection{Interpreting the Results}
Once the map is complete, students can explore spatial clustering and discuss key takeaways:
\begin{itemize}
    \item  The high concentration of cholera deaths around a single water pump (Broad Street pump) supports Snow's hypothesis.
    \item Some deaths occur further away, leading to discussions of outliers, confounding factors and alternative explanations - in interviewing Soho residents, John Snow found that some of those who died from cholera were known to prefer to collect water from the Broad Street pump even though it was not their nearest pump \citep{brody2000map}.
    \item Additional analytical techniques, such as kernel density estimation (KDE) or spatial point pattern analysis, can be introduced to further explore the data. 
\end{itemize}

For example, students could calculate the distance from each cholera case to the nearest water pump to quantify spatial patterns. 

\begin{verbatim}
data(streets)
streets <- streets %>% mutate(across(everything(), as.numeric))
  
# Convert to sf with LINESTRING
streets <- streets %>%
  rowwise() %>%
  mutate(geometry = st_sfc(st_linestring(matrix(c(start_coord_east,
                                start_coord_north, 
                                end_coord_east,      
                                end_coord_north), 
                                ncol = 2, 
                                byrow = TRUE)))) %>%
  ungroup() %>%
  st_as_sf(crs = 27700) 
  

cholera_cases <- st_as_sf(cholera_cases, 
                        coords = c("Easting", "Northing"), 
                        crs = 27700)
pump_locations <- st_as_sf(pump_locations, 
                        coords = c("Easting", "Northing"), 
                        crs = 27700)

# Calculate the nearest pump for each cholera death
cholera_cases <- cholera_cases %>%
  mutate(
   nearest_pump = st_nearest_feature(geometry, pump_locations$geometry,
   dist = as.numeric(min(st_distance(geometry,
                         st_geometry(pump_locations))))))


# Create lines connecting each death to its nearest pump
connections <- cholera_cases %>%
  rowwise() %>%
  mutate(line = st_sfc(st_cast(st_union(geometry, 
  pump_locations$geometry[nearest_pump]), "LINESTRING"))) %>%
  ungroup() %>% st_as_sf()

connections$pump_id<-as.factor(connections$nearest_pump)

ggplot() +
  geom_sf(data = connections, aes(geometry = line, color = pump_id), size = 0.5) + 
  geom_sf(data = cholera_cases, color="black", size = 0.5, alpha = 0.8) +  
  geom_sf(data = pump_locations, color = "red", size = 3, pch = 8) + 
  geom_sf(data = streets, alpha = 0.2)+
  theme_minimal() +
  labs(color = "Pump ID")  
\end{verbatim}

\begin{figure}
    \centering
    \includegraphics[width=\linewidth]{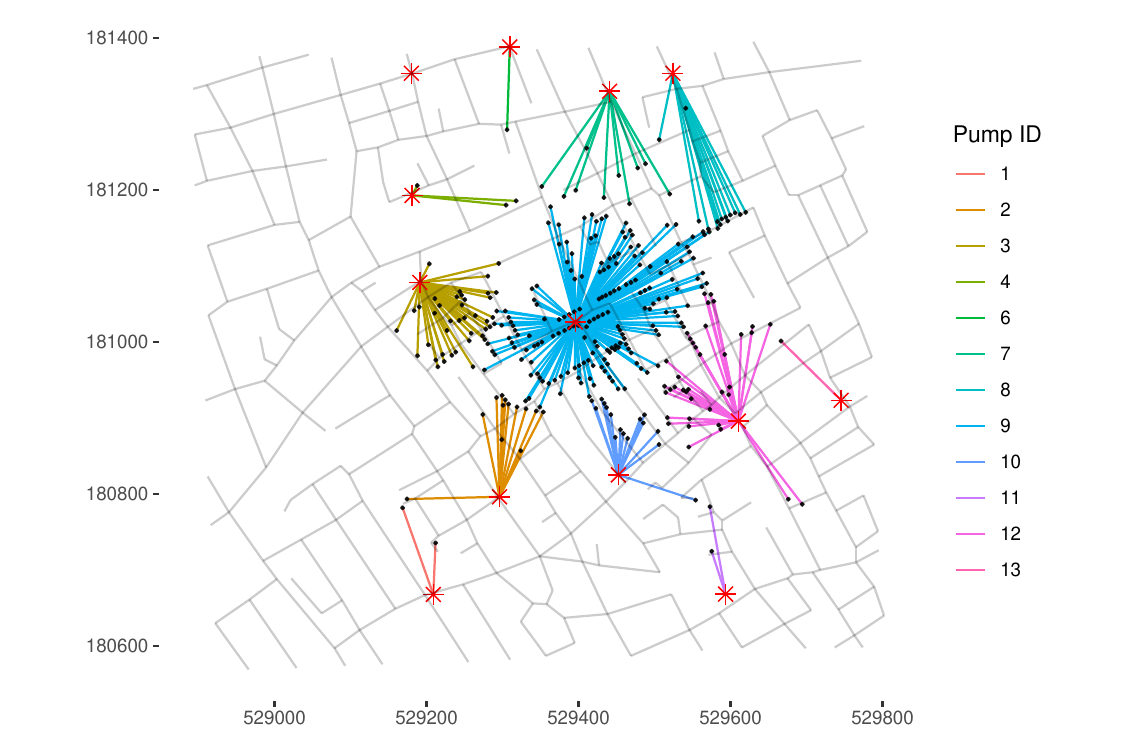}
    \caption{Cholera cases and water pumps overlaid on the Soho street network, with each case connected to its nearest pump. Lines are colour-coded by pump, illustrating potential zones of influence and supporting John Snow’s hypothesis of waterborne transmission.}
    \label{fig:plot4}
\end{figure}
This type of analysis introduces students to spatial statistics, reinforcing the idea that data visualisation is just the first step in an interactive analytical process.

\subsection{Extending the Analysis}
Once students have recreated Snow's map, they can modify the visualisation and extend their analysis in several ways:

\begin{itemize}
    \item Comparing different visualisation types (e.g., heatmaps versus point maps).
    \item Adding interactivity using shiny or leaflet to explore data dynamically.
    \item Applying similar methods to modern epidemiological datasets (e.g., COVID-19 case data).
    \item Performing predicting modelling using machine learning to identify potential hotspots based on simulated data.
\end{itemize}

These extensions help reinforce the practical applications of spatial statistics in real-world scenarios. 

Recreating John Snow's cholera map using R is an effective way to teach students the power of data visualisation in statistical reasoning. This exercise introduce key concepts in exploratory data analysis, geospatial statistics and epidemiology, while also building practical computational skills. By engaging with historical datasets in a modern statistical framework, students gain a deeper appreciation of how data visualisation informs decision-making, both past and present.

\section{Pedagogical Considerations and Applications}
The recreation of John Snow's cholera map using R offers a rich opportunity for educators to introduce students to key concepts in data visualisation, spatial statistics and epidemiology. However, to maximise the education impact of this exercise, instructors should consider how best to structure the activity, tailor it to different levels of student experience, and connect it to broader statistical and data science principles.

In this section, I outline practical strategies for incorporating this case study into statistics education, discuss common challenges and solutions, and suggest extensions that can help students deepen their understanding of spatial analysis and epidemiological modelling. 

\subsection{Integrating the Activity into a Statistics Curriculum}
The John Snow cholera case study can be integrated into a variety of statistics and data science courses, including:

\begin{itemize}
    \item Introductory Statistics: Teaching basic data visualisation, exploratory data analysis, and the importance of graphical methods in discovering patterns.
    \item Applied Data Science: Demonstrating the power of geospatial analysis and reproducible research using R.
    \item Public Health and Epidemiology: Connecting statistical reasoning to real-world disease modelling and outbreak investigation.
    \item History of Statistics: Providing a historical perspective on data-driven decision-making and how early statisticians shaped modern scientific inquiry.
\end{itemize}

Following the guidance of \cite{cobb1992teaching}, instructors may choose to structure the activity as:
\begin{itemize}
    \item A guided tutorial, where students follow a step-by-step replication of John Snow's analysis.
    \item An open ended project, where students experiment with different visualisation techniques and alternative datasets.
    \item A discussion-based case study, using Snow's map as a starting point for critical thinking about data ethics, causality and statistical inference.
\end{itemize}

\subsection{Common Challenges and How to Address Them}
While the John Snow case study is engaging and historically significant, students may encounter several challenges when working with geospatial data in R, including those detailed by \cite{kross2019practitioners}. Below are some potential difficulties and strategies for addressing them.

\begin{table}[ht]
    \centering
    \renewcommand{\arraystretch}{1.5} 
    \begin{tabularx}{\textwidth}{X X}
    \hline
      Challenge   & Solution \\
      \hline
      Students unfamiliar with R  & Provide starter code and structured walk-throughs to build confidence \\
      Geospatial data concepts are new & Begin with simple visualisations (scatterplots) before introducing spatial mapping tools \\
      Data manipulation is complex & Use \texttt{tidyverse} functions to introduce step-by-step data transformation \\
      Students struggle with interpreting spatial patterns & Use interactive tools (shiny, leaflet) to allow dynamic exploration of the data \\
      Linking historical insights to modern applications & Assign follow-up tasks using current datasets (e.g., COVID-19 cases, air pollution data) \\
      \hline
    \end{tabularx}
    \caption{Challenges that may arise in implementing the John Snow case study, and potential solutions.}
    \label{tab:my_label}
\end{table}

By anticipating these difficulties, educators can structure the learning process effectively and ensure that students engage with the material in a meaningful way.

\subsection{Extending the Case Study for Deeper Learning}

Once students have successfully recreated Snow's map, instructors can encourage further exploration through additional activities:
\begin{enumerate}
    \item Data Interpretation and Statistical Inference
    \begin{itemize}
        \item Ask students to quantify the clustering of cholera cases using nearest-neighbour distances or spatial autocorrelation methods
        \item Compare different visualisation methods, such as heatmaps, density plots or contour plots, to illustrate how different graphical techniques highlight patterns
    \end{itemize}
    \item Applying Spatial Analysis to Modern Data
    \begin{itemize}
        \item Extend the methods learned in this case study to modern epidemiological data (e.g., COVID-19 cases, influenza outbreaks).
        \item Use open datasets (e.g., public health, crime data, environmental monitoring) to explore spatial clustering in different contexts.
        \item Compare Snow's approach to modern machine learning methods for disease mapping and risk prediction.
    \end{itemize}
    \item Creating an interactive Shiny App
    \begin{itemize}
        \item Students with programming experience can build an interactive Shiny dashboard that allows users to explore cholera case distributions dynamically.
        \item By incorporating slider controls, hover-over tooltips, and filtering options, students can experiment with making data visualisations more engaging and accessible.
    \end{itemize}
    \item Ethical Discussions on Data-Driven Decision-Making
    \begin{itemize}
        \item Snow's work ultimately influenced public health policy by leading to improvements in sanitation infrastructure.
        \item Ask students to discuss modern ethical considerations in epidemiological modelling, such as data privacy, bias in disease surveillance, and the use of AI for outbreak prediction.
    \end{itemize}
\end{enumerate}

\subsection{Reflections on the Educational Value of Historical Case Studies}

Using historical examples like John Snow's cholera map provides students with more than just technical skills, it also helps them develop an appreciation for the role of statisticians in shaping scientific knowledge and public policy. This case study reinforces several broader lessons.

\begin{itemize}
    \item The importance of visualisation in statistical discovery: Snow's work illustrates that seeing patterns in data is often the first step in generating hypotheses.
    \item The power of interdisciplinary thinking: Combining spatial data, medical knowledge and statistical reasoning led to a groundbreaking statistical advancement.
    \item The value of open data and reproducibility: By recreating Snow's map in R, students engage in the same investigative process that epidemiologists and data scientists use today.
\end{itemize}

Through hands-on exploration, students not only gain computational and analytical skills, but also develop a deeper understanding of how statistics contributes to real-world problem-solving. This blend of technical knowledge, historical context, and critical thinking makes the John Snow cholera map an exceptional teaching tool for modern statistics education.

\section{Conclusion and Future Directions}
John Snow's cholera map remains an incredibly compelling example of how data visualisation and statistical reasoning can uncover hidden patterns and drive real-world change. By recreating and analysing this historical case study using R, students not only engage with foundational concepts in spatial statistics and epidemiology, but also develop computational skills essential for modern data science.

Through this exercise, learners gain experience in data wrangling and geospatial visualisation using R packages, exploratory data analysis and pattern recognition in a real-world historical dataset, and critical thinking about statistical inference and causality, using a case where evidence led to major public health interventions.

Additionally, this pedagogical approach highlights the interdisciplinary nature of statistical thinking, connecting historical data analysis with contemporary public health and policy applications. The ability to bridge historical insights with modern computational methods fosters a deeper appreciation for the evolving role of statistics in decision-making.

\subsection{Expanding the Approach to Other Datasets}
While John Snow's cholera map is an ideal case study for teaching spatial statistics, the underlying principles can be extended to many other datasets. Future implementations of this approach could include:

\begin{itemize}
    \item Contemporary Public Health Data

Analysing COVID-19 case distributions, vaccination uptake or air pollution data, and comparing different data visualisation techniques to communicate health risks effectively.

    \item Environmental and Climate Data

Mapping extreme weather events and their impact on communities, and exploring spatial patterns in climate-related data, such as temperature anomalies or flood occurrences.

    \item Urban Analytics and Social Science Applications

Using geospatial data to analyse crime rates, transportation networks, or socioeconomic disparities, and applying machine learning techniques to detect spatial clustering in these large datasets.
\end{itemize}

By incorporating these extensions, educators can adapt this historical case study into a versatile framework for teaching modern data science skills.

\subsection{Final Remarks}
The use of historical case studies in teaching statistics provides an engaging, tangible way to illustrate core concepts while also highlighting the societal impact of data-driven decision-making. By revisiting Snow's work through a modern computation lens, not do we only honour one of the earliest applications of spatial statistics, but also equip students with the skills to tackle today's pressing data analysis challenges.

\newpage
\bibliographystyle{apalike}  
\bibliography{references}

\begin{thebibliography}{}

\bibitem[Brody et~al., 2000]{brody2000map}
Brody, H., Rip, M.~R., Vinten-Johansen, P., Paneth, N., and Rachman, S. (2000).
\newblock Map-making and myth-making in broad street: the london cholera epidemic, 1854.
\newblock {\em The Lancet}, 356(9223):64--68.

\bibitem[Brunekreef and Holgate, 2002]{brunekreef2002air}
Brunekreef, B. and Holgate, S.~T. (2002).
\newblock Air pollution and health.
\newblock {\em The lancet}, 360(9341):1233--1242.

\bibitem[Cobb, 1992]{cobb1992teaching}
Cobb, G. (1992).
\newblock Teaching statistics.
\newblock {\em Heeding the call for change: Suggestions for curricular action}, 22:3--43.

\bibitem[Cromley and McLafferty, 2011]{cromley2011gis}
Cromley, E.~K. and McLafferty, S.~L. (2011).
\newblock {\em GIS and public health}.
\newblock Guilford Press.

\bibitem[Franch-Pardo et~al., 2020]{franch2020spatial}
Franch-Pardo, I., Napoletano, B.~M., Rosete-Verges, F., and Billa, L. (2020).
\newblock Spatial analysis and gis in the study of covid-19. a review.
\newblock {\em Science of the total environment}, 739:140033.

\bibitem[Friendly and Denis, 2001]{friendly2001milestones}
Friendly, M. and Denis, D.~J. (2001).
\newblock Milestones in the history of thematic cartography, statistical graphics, and data visualization.
\newblock {\em URL http://www. datavis. ca/milestones}, 32:13.

\bibitem[Kross and Guo, 2019]{kross2019practitioners}
Kross, S. and Guo, P.~J. (2019).
\newblock Practitioners teaching data science in industry and academia: Expectations, workflows, and challenges.
\newblock In {\em Proceedings of the 2019 CHI conference on human factors in computing systems}, pages 1--14.

\bibitem[Mimnagh, 2025]{snowdata}
Mimnagh, N. (2025).
\newblock {\em SnowData: Historical Data from John Snow's 1854 Cholera Outbreak Map}.
\newblock R package version 1.0.0.

\bibitem[Nolan and Perrett, 2016]{nolan2016teaching}
Nolan, D. and Perrett, J. (2016).
\newblock Teaching and learning data visualization: Ideas and assignments.
\newblock {\em The American Statistician}, 70(3):260--269.

\bibitem[Pebesma, 2018]{sf2}
Pebesma, E. (2018).
\newblock {Simple Features for R: Standardized Support for Spatial Vector Data}.
\newblock {\em {The R Journal}}, 10(1):439--446.

\bibitem[Pebesma and Bivand, 2023]{sf1}
Pebesma, E. and Bivand, R. (2023).
\newblock {\em {Spatial Data Science: With applications in R}}.
\newblock {Chapman and Hall/CRC}.

\bibitem[Snow, 1856]{snow1856mode}
Snow, J. (1856).
\newblock On the mode of communication of cholera.
\newblock {\em Edinburgh medical journal}, 1(7):668.

\bibitem[Tufte and Robins, 1997]{tufte1997visual}
Tufte, E.~R. and Robins, D. (1997).
\newblock {\em Visual explanations}.
\newblock Graphics Cheshire, CT.

\bibitem[Tukey, 1972]{tukey1972some}
Tukey, J.~W. (1972).
\newblock Some graphic and semigraphic displays.
\newblock {\em Statistical papers in honor of George W. Snedecor}, 5:293--316.

\bibitem[Tukey et~al., 1977]{tukey1977exploratory}
Tukey, J.~W. et~al. (1977).
\newblock {\em Exploratory data analysis}, volume~2.
\newblock Springer.

\bibitem[Unwin, 2018]{unwin2018graphical}
Unwin, A. (2018).
\newblock {\em Graphical data analysis with R}.
\newblock Chapman and Hall/CRC.

\bibitem[Unwin, 2020]{unwin2020data}
Unwin, A. (2020).
\newblock Why is data visualization important? what is important in data visualization.
\newblock {\em Harvard Data Science Review}, 2(1):1.

\bibitem[Wickham, 2016]{ggplot2}
Wickham, H. (2016).
\newblock {\em ggplot2: Elegant Graphics for Data Analysis}.
\newblock Springer-Verlag New York.

\bibitem[Wickham et~al., 2019]{tidyverse}
Wickham, H., Averick, M., Bryan, J., Chang, W., McGowan, L.~D., François, R., Grolemund, G., Hayes, A., Henry, L., Hester, J., Kuhn, M., Pedersen, T.~L., Miller, E., Bache, S.~M., Müller, K., Ooms, J., Robinson, D., Seidel, D.~P., Spinu, V., Takahashi, K., Vaughan, D., Wilke, C., Woo, K., and Yutani, H. (2019).
\newblock Welcome to the {tidyverse}.
\newblock {\em Journal of Open Source Software}, 4(43):1686.

\bibitem[Wickham et~al., 2023]{dplyr}
Wickham, H., François, R., Henry, L., Müller, K., and Vaughan, D. (2023).
\newblock {\em dplyr: A Grammar of Data Manipulation}.
\newblock R package version 1.1.4, https://github.com/tidyverse/dplyr.

\bibitem[Zhang and Schwartz, 2020]{zhang2020spatial}
Zhang, C.~H. and Schwartz, G.~G. (2020).
\newblock Spatial disparities in coronavirus incidence and mortality in the united states: an ecological analysis as of may 2020.
\newblock {\em The Journal of Rural Health}, 36(3):433--445.

\end{thebibliography}
\end{document}